\documentclass[conference]{IEEEtran}
\usepackage{silence}
\IEEEoverridecommandlockouts
\usepackage{cite}
\usepackage{amsmath,amssymb,amsfonts}
\usepackage{algorithmic}
\usepackage{graphicx}
\usepackage{textcomp}
\usepackage{xcolor}
\usepackage{hyperref}

\usepackage{caption}
\usepackage{subcaption}

\begin{document}

\title{Pooling in Graph Convolutional Neural Networks\\
\thanks{This material is based upon work partially funded and supported by the Department of Defense under Contract No. FA8702-15-D-0002 with Carnegie Mellon University for the operation of the Software Engineering Institute, a federally funded research and development center. This work is also partially supported by NSF grants CCF~1837607 and CCN 1513936.}
}

\makeatletter
\newcommand{\linebreakand}{%
  \end{@IEEEauthorhalign}
  \hfill\mbox{}\par
  \mbox{}\hfill\begin{@IEEEauthorhalign}
}
\makeatother

\author{\IEEEauthorblockN{Mark Cheung}
\IEEEauthorblockA{\textit{Electrical and Computer Engineering} \\
\textit{Carnegie Mellon University}\\
Pittsburgh, PA, USA \\
markcheu@andrew.cmu.edu}
\and
\IEEEauthorblockN{John Shi}
\IEEEauthorblockA{\textit{Electrical and Computer Engineering} \\
\textit{Carnegie Mellon University}\\
Pittsburgh, PA, USA \\
jshi3@andrew.cmu.edu}
\and
\IEEEauthorblockN{Lavender Jiang}
\IEEEauthorblockA{\textit{Electrical and Computer Engineering} \\
\textit{Carnegie Mellon University}\\
Pittsburgh, PA, USA \\
yaoj@andrew.cmu.edu}
\linebreakand
\IEEEauthorblockN{Oren Wright}
\IEEEauthorblockA{\textit{Software Engineering Institute} \\
\textit{Carnegie Mellon University}\\
Pittsburgh, PA, USA \\
owright@sei.cmu.edu}
\and
\IEEEauthorblockN{Jos\'{e} M.F. Moura}
\IEEEauthorblockA{\textit{Electrical and Computer Engineering} \\
\textit{Carnegie Mellon University}\\
Pittsburgh, PA, USA \\
moura@andrew.cmu.edu}
}

\maketitle

\begin{abstract}
Graph convolutional neural networks (GCNNs) are a powerful extension of deep learning techniques to graph-structured data problems. We empirically evaluate several pooling methods for GCNNs, and combinations of those graph pooling methods with three different architectures: GCN, TAGCN, and GraphSAGE. We confirm that graph pooling, especially DiffPool, improves classification accuracy on popular graph classification datasets and find that, on average, TAGCN achieves comparable or better accuracy than GCN and GraphSAGE, particularly for datasets with larger and sparser graph structures.
\end{abstract}


\section{Introduction}
Over the past decade, deep learning techniques such as convolutional neural networks (CNNs) have transformed fields like computer vision and other Euclidean data domains (i.e., domains in which data have a uniform, grid-like structure). Many important domains, however, are comprised of non-Euclidean data (i.e., data have irregular relationships that require mathematical concepts like graphs or manifolds to explicitly model). Such domains include social networks, sensor feeds, web traffic, supply chains, and biological systems. As these data grow in size and complexity, deep learning seems to recommend itself as a tool for classification and pattern recognition, but conventional deep learning approaches are often sharply limited when data lack a Euclidean structure to exploit. There are ongoing efforts to extend deep learning to these non-Euclidean domains, and such techniques have been dubbed \textit{geometric deep learning} \cite{bronstein_geometric_2017}.

In parallel with advances in geometric deep learning are advances in graph signal processing (GSP) \cite{sandryhaila2013,ortega_graph_2017}. Research in GSP attempts to generalize classical signal processing theory for irregular data defined on graphs. One attraction of GSP is that it provides a unified mathematical framework through which to view the spectral and vertex domains of a graph. Concepts like frequency or smoothness, which can be understood intuitively in classical signal processing, can be explicitly defined for data on graphs.

Graph convolutional neural networks (GCNNs), an extension of CNNs to graph-structured data, were first implemented with concepts from spectral graph theory \cite{bruna_spectral_2013}, and methods based on the spectral approach have since been refined and expanded \cite{defferrard_convolutional_2016, kipf17}. Reference \cite{du2017} proposes the topology adaptive graph convolutional network (TAGCN) that defines graph convolution directly in the vertex domain as multiplication by polynomials of the graph adjacency matrix. This is consistent with the concept of convolution in graph signal processing \cite{sandryhaila2013}. TAGCN designs a set of fixed-size learnable filters whose topologies are adaptive to the topology of the graph as the filters scan the graph to perform convolution, see also \cite{du2018,shi2018}. Other implementations, such as GraphSAGE \cite{hamilton2017} and graph attention networks (GATs) \cite{Velickovic2018}, are also defined directly in the vertex domain of the graph and apply a learned, convolution-like aggregation function.

An important operation in conventional CNNs is pooling, a nonlinear downsampling operation. Pooling layers in a CNN shrink the number of dimensions of the feature representation, thereby reducing the computation cost, memory footprint, and number of learned parameters. As a result, pooling allows for deeper networks in practice and can help control overfitting. Additionally, pooling has translation invariance properties that are desirable in many applications. Recently, the use of pooling in CNNs has come into question, but it remains popular.

Just as convolution and convolution-like methods have been proposed to create graph convolutional layers in GCNNs, several methods have been proposed in order to perform pooling with GCNNs \cite{ying2018}, \cite{gao2019}, \cite{zhang2018}. Unlike convolution, which has been derived in GSP \cite{sandryhaila2013}, pooling has not been rigorously defined. Therefore, the current generation of pooling methods are based on ad hoc rather than systematic approaches. They nonetheless have shown improved accuracy on popular graph classification datasets.

In this paper, we perform experiments on graph classification datasets, conditionally on graph convolution and graph pooling in GCNNs. This is a supervised learning task in which previously unseen graphs are classified based on labeled graphs. This task is analogous to image classification. Like with CNNs and image classification, tools like pooling layers are important for constructing high-level representations from node-level information.

The paper is divided as follows: we first present the background and related work in section \ref{sec:relatedwork}. Section \ref{sec:proposedmethod} provides our proposed approach. In section \ref{sec:experiments}, we discuss the datasets used and present the results and analysis. Finally, we conclude the paper in section \ref{sec:conclusion}.

\subsection{Graph Signal Processing Perspective}
The convolutional and pooling operator in graph neural network have a theoretical foundation in GSP.
GSP \cite{sandryhaila2013} extends traditional discrete signal processing to graph signals, signals that are indexed by the nodes in a graph. 

Let $G(A) = (V,E)$ be a graph with adjacency matrix $A \in \mathbb{C}^{N\times N}$, where $V$ is the set of $N$ nodes and a nonzero entry $[A]_{ij}$ denotes a directed edge $e_{ij} \in E$ from node $i$ to node $j$.

$\textbf{s}: \mathcal{V} \rightarrow \mathcal{S}$ on $G$ is a graph signal where $\mathcal{S}$ is the signal space over the nodes of $G$ and $\mathcal{S} \subseteq \mathbb{C}^N$. $\textbf{s} = 
(s_1,s_2,...,s_N) \in \mathbb{C}^N$, and $s_i$ represents a measurement at node $v_i \in \mathcal{V}$.

The heart of GCNNs is applying convolutional filters to graph signals. In GSP, convolution is a matrix-vector multiplication of a polynomial of the adjacency matrix $P(A)$ and the graph signal $x$. This definition is used to create the graph convolutional layer in GCNNs.

The GSP literature includes \cite{chen2015} and \cite{anis2016}. In \cite{chen2015} and \cite{anis2016}, several sampling set selection and sampling methods are proposed. The pooling methods explored herein are not based specifically on these sampling methods, but we observe that there is a relationship between sampling in GSP and pooling in GCNNs. Both reduce the number of values in the signal and can reduce the number of nodes in the graph. The key difference is that, in sampling, we focus on how to recover the original signal given the sampled signal. However, recoverability is not required in pooling algorithms in GCNNs.

\section{Related Work}
In this section, we describe the infrastructure for graph convolutional and pooling layers and the related literature.

\label{sec:relatedwork}
\subsection{Graph Convolutional Layer}

We concentrate on three implementations of GCNNs, derived from different definitions of graph convolution: graph convolutional networks (GCNs) \cite{kipf17}, GraphSAGE \cite{hamilton2017}, and topology-adaptive graph convolutional networks (TAGCNs) \cite{du2017,shi2018}.

In GCN \cite{kipf17}, given a graph signal $X^{(0)} \in \mathbb{R}^{n \times c}$ (where $0$ denotes the input layer, $n$ is the number of nodes, and $c$ is the number of features/input channels) and a graph structure $A \in \mathbb{R}^{n \times n}$, a graph convolutional layer is defined as follows:
\begin{equation}
\label{eq:gcn}
    X^{(l+1)}=\sigma ( \widehat{D}^{-\frac{1}{2}}\widehat{A}\widehat{D}^{-\frac{1}{2}}  X^{(l)} W^{(l)})
\end{equation}
where $\widehat{A}=A+I_n$, $\widehat{D}_{ii}=\sum_j \widehat{A}_{ij}$, $W^{(l)} \in \mathbb{R}^{c \times c'}$ is the trainable weight matrix, $\sigma$ is the nonlinear activation function, and $c'$ is the number of output channels. $l=0$ for the first layer, and we can propagate the graph signal through additional layers in the network. This approach is based on a first-order approximation of localized spectral filters on graphs \cite{HAMMOND2011129}.

In GraphSAGE \cite{hamilton2017}, graph convolution is defined as follows, for each node $v$ of $X$:
\begin{equation}
\label{eq:graphsage}
    X_v^{(l+1)}=\sigma (W^{(l)} \cdot f_k\left(X_v^{(l)},\left\{X_u^{(l)}, \forall u \in S_{\mathcal{N}(v)} \right\}\right)
\end{equation}
where $f_k$ is an aggregator function (e.g., sum, mean, or max), and $S_{\mathcal{N}(v)}$ is a random sample of the node $v$'s neighbors.

In TAGCN \cite{du2017}, it is defined as follows:
\begin{equation}
     X^{(l+1)} = \sigma \left(\sum_{i=0}^N  (D^{-\frac{1}{2}}A D^{-\frac{1}{2}})^i  X^{(l)} W_i^{(l)}\right)
\end{equation}
where  $D_{ii}=\sum_j A_{ij}$ and $W_{0 \dots n}^{(l)} \in \mathbb{R}^{c \times c'}$.

\subsection{Graph Pooling Layer}

Similar to graph convolution, graph pooling is inspired by pooling in CNNs. In addition to static pooling methods \cite{Luzhnica2019,boykov2006}, various differentiable methods have been proposed.

Using the same notation as \eqref{eq:gcn}, a graph pooling operator should yield a new signal $x' \in \mathbb R^{n' \times c}$ and adjacency matrix $A' \in \mathbb R^{n' \times n'}$, usually with $n \geq n'$. See Fig. \ref{fig:graph_pooling} for an example. 

\begin{figure}
	\includegraphics[width=\linewidth]{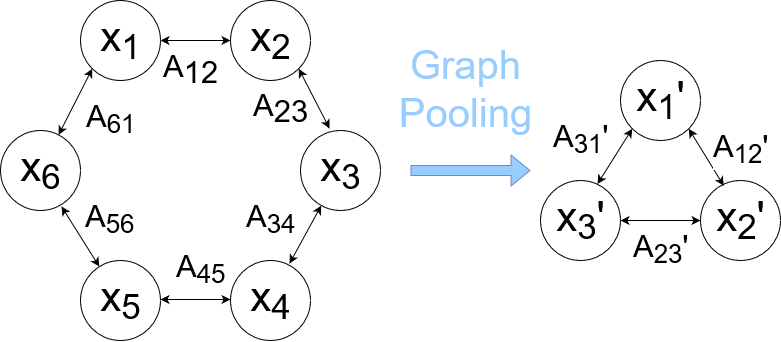}
	\caption{Graph pooling, yielding a new signal and Adjacency matrix}
	\label{fig:graph_pooling}
\end{figure}

An important benefit of graph pooling is the hierarchical representation of data and structure. Otherwise, global patterns in the data are usually not considered until the final aggregation layer of a network. Below we describe four recent graph pooling algorithms.

\subsubsection{Sort Pooling}
Sort Pooling (SortPool) \cite{zhang2018} operates after the last graph convolution layer. Instead of summing or averaging features, SortPool arranges the vertices in a consistent order and outputs a representation with a fixed set, so that further training using CNN can be done.

The vertices are sorted based on their structural roles within the graph. Using the connection between graph convolution and the Weisfeiler-Lehman subtree kernel \cite{Shervashidze2011}, SortPool sorts the node features of the last layer individually, then sorts in descending order based on the layer before, and finally selects the top $k$ nodes.

\subsubsection{Differentiable Pooling}
Differentiable Pooling (DiffPool) \cite{ying2018} is a differentiable graph pooling module that learns hierarchical representations of the graphs by aggregating nodes through several pooling layers. It uses a learned assignment matrix $S \in \mathbb{R}^{n_1 \times n_2}$ and updates the graph signal and topology as follows:

\begin{align}
    X^{(l+1)}&=S^{(l)^T}Z^{(l)} \\
    A^{(l+1)}&=S^{(l)^T}A^{(l)}S^{(l)}
\end{align}
where $Z^{(l)} \in \mathbb{R}^{n_1 \times n_1}$ is the GraphSAGE \cite{hamilton2017} operator with the mean aggregator. DiffPool achieves significantly better prediction accuracy than GraphSAGE, SortPool, and certain kernel methods, especially when global features are important for classification \cite{ying2018}.

\subsubsection{Top-k Pooling}
Top-k Pool \cite{gao2019} pools using a trainable projection vector $p^{(l)} \in \mathbb{R}^{f}$ and select the top-k indices of the projection $X^{(l)} p^{(l)}$ and the corresponding edges in $A^{(l)}$.

% \begin{align}
%     y &= X^l p^l / \norm{p^l} \\
%     \mathrm{idx} &=  \mathrm{rank}(y,k) \\
%     \Tilde{y} &= \mathrm{sigmoid}(y(\mathrm{idx}))\\
%     \Tilde{X}^l&=X^l(\mathrm{idx},:) \\
%     A^{l+1}&=A^{l}(\mathrm{idx},\mathrm{idx}) \\
%     X^{l+1}&=\Tilde{X}^l \odot (\Tilde{y}\mathbf{1}^T_C)
% \end{align}
% where $k$ is the hyperparameter, the number of nodes to pool, $\mathrm{rank}$ select the largest $k$ nodes in $y$.

Top-k pool is inspired by encoder-decoder architectures like U-Nets. In addition to the Top-k pool operation, there is also an Unpool operation that reverses the process. These two combined create the encoder-decoder model on graph, known as the graph U-Nets \cite{gao2019}. Reference \cite{gao2019} shows that Top-k pool with the U-net structure performs better than DiffPool, but we will show if it works well standalone vs. other pooling algorithms.

\subsubsection{Self-Attention Graph Pooling}
Self-Attention Graph Pooling (SagPool) \cite{gao2019} uses an attention mechanism to select the important nodes:
\begin{align}
    y &=\mathrm{GNN}(X^{(l)},A^{(l)}) \\
    i &=\mathrm{top}_k(y) \\
    X^{(l+1)} &= (X^{(l)} \odot \mathrm{tanh}(y))_i \\
    A^{(l+1)} &= A_{i,i}^{(l)}
\end{align}
The attention score is calculated from GCN and the top $k$ nodes are selected from it. Since graph convolution is used to obtain the self-attention score, SagPool uses both the graph features and structure  \cite{gao2019}. Reference \cite{gao2019} shows that SAGPool performs better than DiffPool and Top-k Pool across some biochemical datasets.

\section{Proposed Method}
\label{sec:proposedmethod}
We first compare GCN, GraphSAGE, and TAGCN for graph classification across four benchmark datasets. We then investigate how pooling affects these results, by combining the different convolutional architectures with the four pooling techniques described above, i.e., SortPool \cite{zhang2018}, DiffPool \cite{ying2018}, Top-k Pool \cite{gao2019}, and SagPool \cite{lee2019}. In each instance, the pooling method is paired with GCN or GraphSAGE (determined by that used in each pooling paper), and compared with the pooling method paired with TAGCN.

\section{Experiments}
\label{sec:experiments}
\subsection{Datasets} 
To evaluate the efficacy of the different methods, we apply our methods on real-world graph kernel benchmarks. See Table \ref{table:dataset} for the properties of these datasets. We evaluate our methods on bioinfomatics datasets and social network datassets. Both MUTAG and Proteins datasets are bioinformatics data. MUTAG \cite{mutag} is a dataset consisting of chemical compounds represented by graphs. The task is to predict whether the chemical compound is mutagenic. Proteins \cite{pt2005} is a dataset consisting of  proteins represented by graphs. The objective is to predict whether a protein functions as an enzyme. In both of the datasets, the nodes are  structure elements, and two nodes are connected if there is a chemical bond between the structure elements represented by the nodes.

For social network datasets, we chose IMDB-Binary and Reddit-Binary. IMDB-Binary \cite{Yanardag2015} is a set of graphs corresponding to ego-networks of actors and actresses. An edge is drawn between two actors if they were cast in the same movie. The task is to predict whether a movie is romance or action. In Reddit-Binary \cite{Yanardag2015}, each graph corresponds to an online discussion thread. An edge is drawn between two users if one has replied to the other. The task is to predict whether a thread belongs to a discussion forum or a question answering forum. 

% \begin{table}[!htbp]
% \centering
% \caption{Properties of Node Classification Datasets}
% \begin{tabular}{|c|c|c|c|c|c|}
% \hline
%  Dataset &   Nodes   & Label Rate & Connectivity & Features & Classes \\ \hline
% CORA & 2708 & 0.052 & 7e-4       & 1433    & 7 \\ 
%  \hline
% CiteSeer & 3327 & 0.036 & 4e-4       & 3703     & 6   \\ 
% \hline
% Pubmed & 19717 & 0.003 & 1e-4       & 500     & 3   \\ 
% \hline
% \end{tabular}
% \label{table:fsdataset}
% \end{table}

\begin{table}[!htbp]
\centering
\caption{Properties of Graph Classification Datasets}
\begin{tabular}{|c|c|c|c|c|c|}
\hline
 Dataset &   Graphs   & Classes & Avg Nodes & Avg Edges\\ \hline
MUTAG & 188 & 2 & 17.7       & 38.9    \\ 
 \hline
Proteins & 1113 & 2 & 39.06       & 72.82      \\ 
\hline
IMDB-Binary & 1000 & 2 & 19.77       & 96.53      \\ 
\hline
Reddit-Binary & 2000 & 2 & 429.63       & 497.75    \\ 
\hline
\end{tabular}
\label{table:dataset}
\end{table}

% \begin{table}[!htbp]
% \centering
% \caption{Parameters of Graph Classification Datasets}
% \begin{tabular}{|c|c|c|}
% \hline
%  Dataset &  Average degree   & Clustering coefficient \\
%  \hline
% MUTAG &{\color{red}TODO} &   \\ 
% \hline
% Proteins & &   \\ 
% \hline
% IMDB-Binbary & &   \\ 
% \hline
% Reddit-Binary & &   \\ 
% \hline
% \end{tabular}
% \label{table:metrics}
% \end{table}

% stats from https://ls11-www.cs.tu-dortmund.de/staff/morris/graphkerneldatasets

\subsection{Network Training}
We perform 5-fold cross-validation to select the hyperparameters from the validation accuracy and estimate the test accuracy. For the baselines, the hyperparameters are the number of graph convolutional layers, number of channels in each layer, dropout rates, pooling rate (number or percentage of nodes to keep), and (for TAGCN) order of polynomial filter. For a fairer comparison, we considered 1-5 layers for TAGCN vs. 1-15 layers for GCN and GraphSAGE when using graph polynomial filters of degree~3 (to show that 1 layer of TAGCN with degree $n$ is not $n$ layers of GCN/GraphSAGE). We use cross-entropy loss and ADAM optimization with a starting learning rate of 0.01, a decay factor of 0.5, and a decay step size of 50. Experiments were performed in PyTorch using code from the PyTorch Geometric Library \cite{fey2019}.

\subsection{Results}

% \begin{table}[!htbp]
% \caption{Comparison of GCNN Variant without pooling}
% \begin{tabular}{l|l|l|l|l|}
% \cline{2-5}
%  & MUTAG & Proteins & \begin{tabular}[c]{@{}l@{}}IMDB\\ -Binary\end{tabular} & \begin{tabular}[c]{@{}l@{}}Reddit\\ -Binary\end{tabular} \\ \hline
% \multicolumn{1}{|l|}{GCN}       & $71.8 \pm 3.2$       &  $72.8 \pm 1.7$        &      $ 72.8 \pm 2.3 $     &       $88.2 \pm 1.8 $       \\ \hline
% \multicolumn{1}{|l|}{GraphSAGE} &   $73.9 \pm 3.7$    & $73.6 \pm 2.2$         &       $ 71.8 \pm 4.0 $    &     $ 85.1 \pm 3.4 $        \\ \hline
% \multicolumn{1}{|l|}{TAGCN}     &   $75.1 \pm 8.2$    & $72.4 \pm 2.9 $        &    $73.3 \pm 5.3 $        &  $91.6 \pm 2.6$              \\ \hline
% \end{tabular}
% \label{table:no_pooling}
% \end{table}

% \clearpage
\begin{figure*}[t!]
    \centering
    \begin{subfigure}[t]{0.48\textwidth}
	\includegraphics[width=\linewidth]{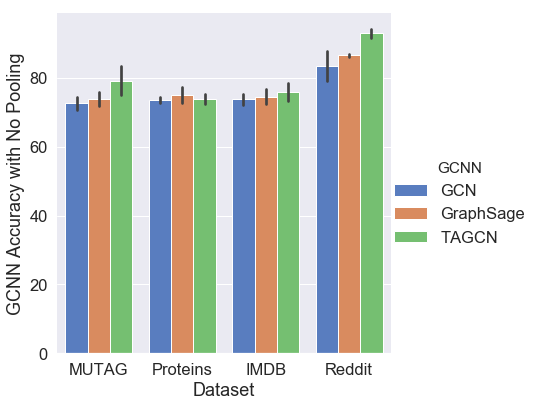}
	\label{fig:no_pooling}
    \end{subfigure}%
        \begin{subfigure}[t]{0.48\textwidth}
    	\includegraphics[width=\linewidth]{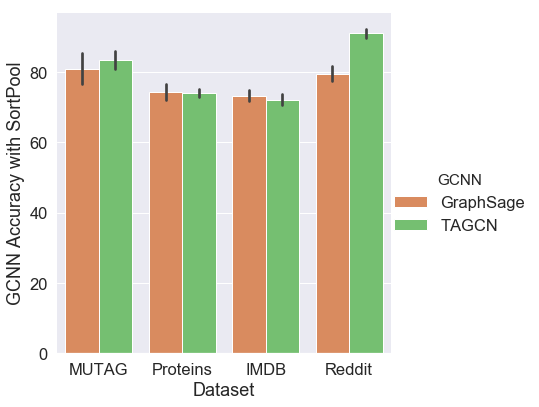}
	\label{fig:sortPool}
    \end{subfigure}
    \begin{subfigure}[t]{0.48\textwidth}
	\includegraphics[width=\linewidth]{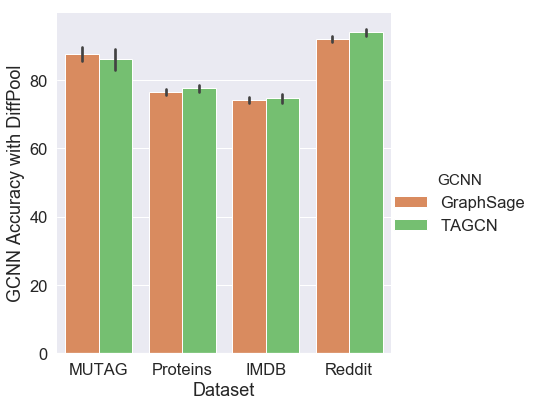}
	\label{fig:diffPool}
    \end{subfigure}
	 \begin{subfigure}[t]{0.48\textwidth}
	\includegraphics[width=\linewidth]{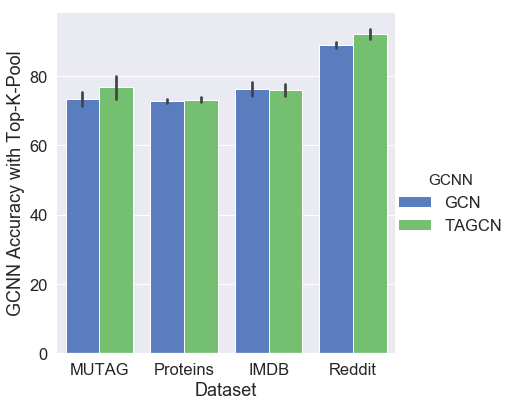}
	\label{fig:TopKPool}
    \end{subfigure}
    \begin{subfigure}[t]{0.48\textwidth}
	\includegraphics[width=\linewidth]{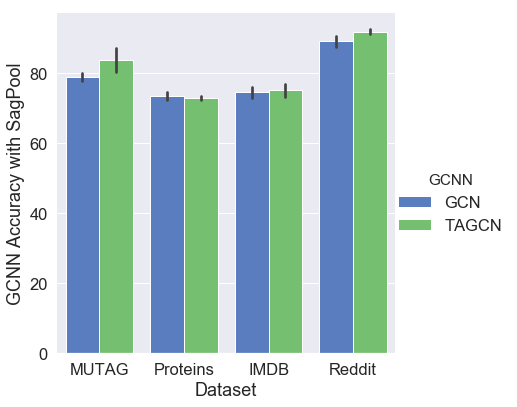}
	\label{fig:sagPool}
    \end{subfigure}
	
	\caption{Comparison of graph classification accuracies of GCNN Variant (TAGCN, GCN, GraphSAGE) for no pooling, SortPool, DiffPool, Top-k Pool, and SagPool across 4 datasets (MUTAG, Proteins, IMDB-Binary, Reddit-Binary)}
	\label{fig:pooling}
\end{figure*}

Fig. \ref{fig:pooling} shows the results of GCNN variants with no pooling, DiffPool, SagPool, SortPool, and Top-K Pool. The green, orange, and blue bars are the means of the cross-validated accuracy and the smaller black error bars are their standard deviations.

\subsubsection{Graph Convolution Comparison}
In general, TAGCN performs better than GCN and GraphSAGE on the four graph classification benchmarks. However, due to the increase in complexity, TAGCN has high variance, especially denser graph structures. TAGCN performs better as graphs become less sparse, i.e., as average degree increases.

We also showed empirically that simply increasing number of layers in GCN and GraphSAGE is not analogous to increasing the order of the polynomial filter in TAGCN. We attribute the the advantage of TAGCN mainly to: 1) Passing a residual connection of the graph signal, and 2) Having weights associated with each polynomial of the adjacency matrix. In comparison, GCN and GraphSAGE do not improve much after five layers, perhaps also suffering from oversmoothing.

\subsubsection{Graph Convolution and Graph Pooling Comparison}
Among the pooling algorithms, DiffPool generally performs the best. SagPool and SortPool perform better for MUTAG and Proteins, but similar or worse for IMDB-Binary and Reddit-Binary. Top-k pool performs poorly, suggesting that it requires the auto-encoder structure to perform better. In general, only Diffpool is consistently better than no pooling. 

The results for graph convolution apply to graph pooling with graph convolution. TAGCN with pooling generally performs better than GCN and GraphSAGE with pooling and more prone to overfitting, likely due to the same reasons.

\section{Conclusion}
\label{sec:conclusion}
On average, TAGCN generally performs well against GCN and GraphSAGE on graph classification datasets with and without pooling for sparser and larger graphs. We also find that DiffPool generally outperforms the other pooling methods evaluated. For future work, we would like to develop a better theoretical understanding of GCNNs, by studying different problems like oversmoothing and the design of different parameters.

\bibliographystyle{IEEEtran}

\bibliography{ref}

\end{document}